\documentclass[aps,prb,preprint,groupedaddress,showpacs,showkeys]{revtex4}
\usepackage{graphicx}
\usepackage[english]{babel}
\newcommand{\etal}{\emph{et al.}}

\begin{document}


\title{Exciton bimolecular annihilation dynamics in supramolecular nanostructures of conjugated oligomers}

\author{Cl\'{e}ment Daniel}
\author{Laura M. Herz}
\author{Carlos Silva}
\email[Corresponding author. ]{cs271@cam.ac.uk}
\homepage[]{http://www-oe.phy.cam.ac.uk/fast/index.htm}
\affiliation{Cavendish Laboratory, University of Cambridge, Madingley Road, Cambridge CB3 0HE, United Kingdom}
\author{Freek J. M. Hoeben}
\author{Pascal Jonkheijm}
\author{Albertus P. H. J. Schenning}
\author{E. W. Meijer}
\affiliation{Laboratory of Macromolecular and Organic Chemistry, Eindhoven University of Technology, P.O. Box 513, 5600 MB Eindhoven, The Netherlands}



\date{\today}

\begin{abstract}
We present femtosecond transient absorption measurements on $\pi$-conjugated supramolecular assemblies in a high pump fluence regime. Oligo(\emph{p}-phenylenevinylene) monofunctionalized with ureido-\emph{s}-triazine (MOPV) self-assembles into chiral stacks in dodecane solution below 75$^{\circ}$C at a concentration of $4\times 10^{-4}$\,M. We observe exciton bimolecular annihilation in MOPV stacks at high excitation fluence, indicated by the fluence-dependent decay of $1^1$B$_{u}$-exciton spectral signatures, and by the sub-linear fluence dependence of time- and wavelength-integrated photoluminescence (PL) intensity. These two characteristics are much less pronounced in MOPV solution where the phase equilibrium is shifted significantly away from supramolecular assembly, slightly below the transition temperature. A mesoscopic rate-equation model is applied to extract the bimolecular annihilation rate constant from the excitation fluence dependence of transient absorption and PL signals. The results demonstrate that the bimolecular annihilation rate is very high with a square-root dependence in time. The exciton annihilation results from a combination of fast exciton diffusion and resonance energy transfer. The supramolecular nanostructures studied here have electronic properties that are intermediate between molecular aggregates and polymeric semiconductors.

\end{abstract}


\maketitle


\section{Introduction}
Organic semiconductor materials are now widely used in optoelectronics devices such as field-effect transistors,\cite{Stutzmann03,Dimitrakopoulos02} light-emitting diodes,\cite{Friend99,Forrest00} and photovoltaic diodes.\cite{Brabec02,Gregg03,Peumans03} Intramolecular functionality ($\pi$-conjugation) can be tailored with synthetic methodologies,\cite{Setayesh01,Ego03,Grimsdale02} but control of intermolecular properties, which are equally important in determining electronic properties (\emph{e.g.} charge transport), is more elusive. Perhaps the most compelling advantage of many organic semiconductors such as conjugated polymers over molecular and inorganic semiconductors is solution processability, making fabrication of elegantly simple device structures possible with techniques such as ink-jet printing.\cite{Sirringhaus00,Kawase01} A natural strategy to achieve three-dimensional control of intermolecular interactions is to exploit molecular self assembly in solution prior to the casting process by means of supramolecular chemistry.\cite{Lehn95} With this approach, molecular building blocks are self-assembled to form well-defined, complex architectures through secondary forces, such as electrophilic/electrophobic interactions. These interactions need to be strong enough to lead to spontaneous self-organization but weak enough so that the process is reversible. Supramolecular assemblies possess polymeric characteristics and are therefore often referred to as "supramolecular polymers". Their macroscopic properties can be tuned such that high carrier mobilities, for example, can be achieved.\cite{Craats99}

Recently, this approach has been applied very successfully to oligo(\emph{p}-phenylenevinylene) derivatives with chiral side-chains and functionalised with ureido-\emph{s}-triazine, a hydrogen-bonding end-group.\cite{Schenning01,Meijer02} It has been shown that these monofunctionalized oligo(\emph{p}-phenylenevinylene) (MOPV) self-organize in chiral stacks in apolar solvents below a transition temperature. Here, the secondary interactions used for the supramolecular assembly are quadruple hydrogen bonds, $\pi$-$\pi$ stacking and solvophobic effects.

Resonance energy transfer is fundamental to describe exciton dynamics in conjugated structures. It has been shown that intermolecular energy transfer is the dominant mechanism in polymer films while intramolecular energy transfer determines exciton dynamics in dilute polymer solutions.\cite{Nguyen00a,silva02,Beljonne02} A quantitative study of intermolecular energy transfer in polymer films is complicated by the intrinsic positional and energy disorder of such systems. MOPV, on the other hand, is a model system to study energy transfer dynamics as it is possible to investigate the effect of intermolecular interactions by comparing optical properties of MOPV in the dissolved phase and the supramolecular assemblies. In supramolecular assemblies, the conjugated segments are closely packed in a fashion similar to conjugated segments in polymeric films but in a well-defined manner. In a previous paper, we have investigated exciton dynamics in MOPV at low excitation fluence with time-resolved photoluminescence spectroscopy and we found them to be very similar to those of polymeric films.\cite{Herz03a} Fast exciton diffusion on chiral supramolecular assemblies was shown to lead to exciton trapping and luminescence depolarization.

It has been documented that at sufficiently high excitation
fluence, exciton dynamics are dominated by bimolecular
annihilation processes. This exciton-exciton interaction has been studied
in several polymeric systems\cite{Kepler96,Klimov97,Vacar97,Maniloff97,Dogariu98,Denton99,Nguyen00,Stevens01,Shimizu01,Silva01} (see Gadermaier~\etal\cite{Gadermaier02b} for a recent review on ultrafast studies of exciton dynamics in polymeric semiconductors). Two
models have been proposed for the mechanisms of exciton-exciton
annihilation and have been applied to various polymeric systems.
The first model assumes the rate-limiting process to be the
diffusion and encounter of excitons, with the rate constant being
time-independent.\cite{Kepler96,Denton99,Shimizu01,Stevens01,Silva01} In the second model,
the rate-limiting process is considered to be a long-range resonance-energy-transfer between the excitons.
The proposed form of the rate constant varies but usually has an explicit time dependence due to the process being
non-Markovian.\cite{Dogariu98,Vacar97,Maniloff97,Nguyen00,Klimov97} These two mechanisms are in competition, and the key parameter that determines the importance of the non-Markovian mechanism is the spectral overlap of the 1$^{1}$B$_{u}$ absorption and emission spectra.\cite{Stevens01} The excited-state absorption spectrum is generally broad and its position is similar in polyphenylene and polyphenylenevinylene materials, but the PL spectral position is more sensitive to chemical composition. As a result, exciton bimolecular interactions in green and red-emitting materials tend to be dominated by long-range resonance effects, while in blue-emitting materials collisional processes tend to dominate bimolecular annihilation dynamics.

In this paper we describe a comprehensive study of exciton bimolecular annihilation in MOPV supramolecular assemblies in dilute solution. Applying femtosecond transient absorption spectroscopy, we show that excitation with pulse fluences above $\sim 50$\,$\mu$J/cm$^{2}$ results in bimolecular annihilation dynamics that dominate on picosecond timescales. This is in analogy to solid films of polymeric semiconductors. Long-range exciton bimolecular interactions determine the annihilation dynamics as has been reported for related alkoxy-substituted polyphenylenevinylene films. These results demonstrate that the electronic properties of solution-self-assembled supramolecular architectures are similar to those of disordered polymeric semiconductors. However, supramolecular electronics allow tailored control of intermolecular electronic interactions, as well as reversibility of the assembly process.

\section{Experimental}

\begin{figure}
\includegraphics{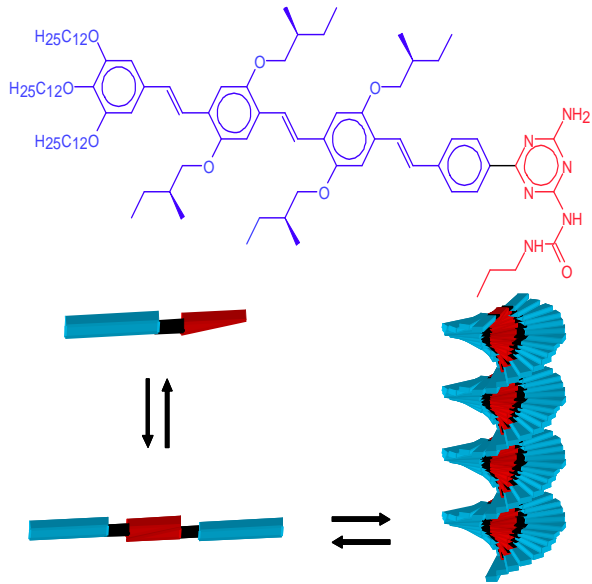}
\caption{\label{fig:MOPV4}Molecular structure of MOPV4. In dodecane, quadrupolar hydrogen bonding of the ureido-\emph{s}-triazine unit leads to formation of dimers, which self-assemble into chiral structures as depicted in the schematic diagram. The supramolecular assembly is thermotropically reversible with a transition temperature of 75$^{\circ}$C at an MOPV4 concentration of $4 \times 10^{-4}$\,M.}
\end{figure}

The synthesis of MOPV4 (see Figure~\ref{fig:MOPV4} for the molecular structure) has been described elsewhere.\cite{Schenning01} The material was dissolved in anhydrous dodecane at a concentration of $4\times10^{-4}$\,M. Measurements were carried in a temperature-controlled absorption cuvette with 1\,mm path length.

The femtosecond transient absorption apparatus consisted of a laser system based on the design of Backus~\etal\cite{Backus98} Briefly, 15-fs pulses were produced by a mode-locked Ti:sapphire oscillator (KMLabs TS pumped by 4.5\,W, 532\,nm output from a Spectra-Physics Millennia V laser). The pulse train was amplified using a chirped-pulse-amplification scheme at a repetition rate of 1\,kHz. The amplifier was a multipass Ti:sapphire amplifier (pumped by the 9.5\,W, 527\,nm output from a Spectra-Physics Evolution X laser). The resulting pulses had a pulse width of 100\,fs, a pulse energy of 400\,$\mu$J and were centered at a wavelength of 800\,nm (1.55\,eV photon energy).

The sample was excited with 3.1\,eV pump pulses produced by frequency doubling the laser fundamental in a 0.5-mm pathlength $\beta$-BBO crystal. The pump power was varied with calibrated neutral density filters and the beam was mechanically chopped at 500\,Hz. The transient absorption of the sample was probed with a single-filament white-light continuum (approximately from 1.2\,eV to 3.0\,eV) produced by focusing a fraction of the fundamental into a sapphire window. Probe pulses were delayed with respect to pump pulses with a computer-controlled optical delay. All the beams were linearly polarized with parallel polarization. The pump and the probe beams were focused on the sample cell to respectively $\sim50$\,$\mu$m and 100\,$\mu$m spots (with an angle between beams of approximatively 20$^{\circ}$). The probe beam and a reference beam (not transmitted through the sample) were dispersed in a 0.25-m spectrometer and detected with a pair of Si photodiodes. Pump-induced variation in transmission ($\Delta T/T$) was extracted using standard lockin techniques. The instrument response function had a full width at half maximum of less than 120\,fs.

We applied time-correlated single photon counting (TCSPC) to measure excited-state lifetimes. The MOPV solution was excited with a pulsed (20\,MHz, 70\,ps FWHM, 407\,nm) diode laser (PicoQuant LDH400). The luminescence was detected with a microchannel plate photomultiplier (Hamamatsu) coupled to a spectrometer and TCSPC electronics (Edinburgh Instruments Lifespec-ps and VTC900 PCI card). A decay curve was taken at 2.21\,eV photon energy, and the decay time was extracted from a mono-exponential fit to the data.

\section{Results and Analysis}

\begin{figure}
\includegraphics{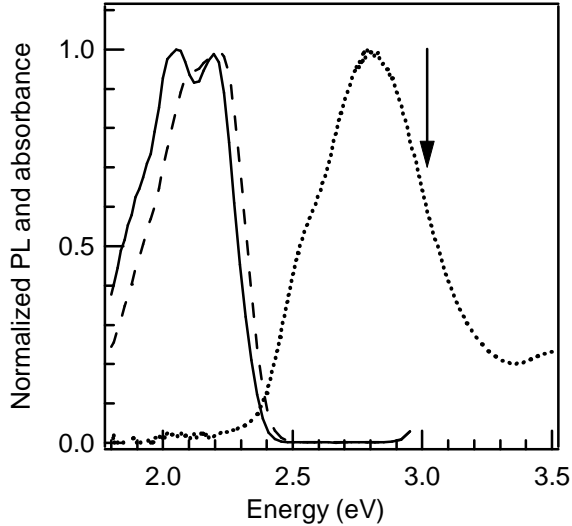}
\caption{\label{fig:PLAbs}Absorption spectrum of MOPV4 at 20$^{\circ}$C (dotted line), PL spectrum at 14$^{\circ}$C (solid line) and PL spectrum at 65$^{\circ}$C (broken line). The black arrow indicates the excitation photon energy in all experiments reported here.}
\end{figure}

The material investigated here is an oligophenynevinylene
consisting of four phenyl rings, monofunctionalized with a
ureido-\emph{s}-triazine (MOPV4) to dimerise by hydrogen bonding (see Figure~\ref{fig:MOPV4}). The absorption spectrum of MOPV4 is shown in Figure~\ref{fig:PLAbs}
together with the photoluminescence (PL) spectrum at 14$^{\circ}$C
and 65$^{\circ}$C for excitation using femtosecond pulses with photon energy just above the peak of the
$\pi$-$\pi^*$ band. A red shift is observed upon cooling the
solution, which has been previously attributed to the formation of
supramolecular assemblies.\cite{Schenning01,Herz03a} Small-angle
neutron scattering experiments showed that the average stack
length is around 160\,nm with a radius around 6\,nm.\cite{Jonkheijm03}
The stack configuration is not known precisely but according to
preliminary quantum-chemical calculations, the oligomer separation
is approximatively 35\,\AA~and the angle of rotation between two
neighboring oligomer is close to 12$^{\circ}$.\cite{BeljonneC}
The intermolecular coupling in the stacks modifies the excitonic
characteristics and leads to the appearance of a red shoulder in
the PL spectrum. Above a certain transition temperature, which depends on
the MOPV4 concentration, the thermodynamic equilibrium shifts from
supramolecular assembly to a dissolved phase.\cite{Schenning01}
Within a relatively narrow range ($\pm \sim 10^{\circ}$C) about the
transition temperature, the positional disorder in the stack
increases and the average assembly length decreases. At the MOPV4
concentration used here, the transition temperature is around
$\sim 75^{\circ}$C. However, the two
PL spectra show that at 65$^{\circ}$C, the red shoulder has
already decreased significantly suggesting that the intermolecular
coupling is much smaller at 65$^{\circ}$C. Therefore, we can use
these two experimental conditions (14$^{\circ}$C and
65$^{\circ}$C) to compare exciton dynamics with and without strong
intermolecular coupling.

\begin{figure}
\includegraphics{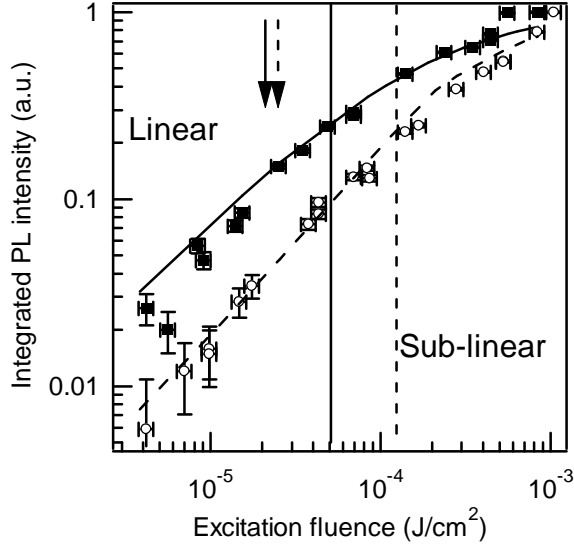}
\caption{\label{fig:PLPow}Time- and wavelength-integrated PL intensity
at 14$^{\circ}$C (filled squares) and at 65$^{\circ}$C (open
circles). The solid and broken lines are the results of fits to
the time- and volume-integrated exciton population density using the model
described in the text. The vertical lines separate linear and
sub-linear regimes for the data taken at 14$^{\circ}$C (solid
line) and 65$^{\circ}$C (broken line). The two arrows point to the
excitation fluence of the data used for the stretched exponential
fits (see Figure \ref{fig:PalowP}).}
\end{figure}

Figure~\ref{fig:PLPow} shows the excitation-fluence dependence of the time- and wavelength-integrated PL intensity of MOPV4 at 14$^{\circ}$C and 65$^{\circ}$C. The dependence is linear with excitation fluence but a sub-linear trend appears at higher fluence. The transition between the two regimes is temperature-dependent. At lower solution temperatures, the PL intensity saturates at significantly lower fluence. Moreover, the deviation from linearity at high excitation fluence is stronger at 14$^{\circ}$C. Such behavior is indicative of a time-integrated exciton population which develops linearly with low pump fluence and sub-linearly with high fluence. This indicates that exciton-exciton annihilation occurs on the supramolecular assemblies at high fluence. At low fluence, on the other hand, we are able to probe a regime where exciton-exciton annihilation dynamics are negligible.

Various OPV derivatives have been previously studied using
femtosecond transient absorption
spectroscopy,\cite{Klimov97,Klimov98,McBranch99,Kraabel00} and
three distinct spectral signatures were identified. The
first is probe-induced stimulated emission (SE) signal overlapped
with the PL spectrum (around 2.21\,eV) while the other responses
are photoinduced absorption (PA) signal around 1.77\,eV and
1.46\,eV. The SE and the 1.46-eV PA were assigned to
1$^1$B$_u$-exciton dynamics while the 1.77-eV PA signal was
assigned to polaron absorption. In the following discussion we
will limit ourselves to the study of the 1$^1$B$_u$-exciton
dynamics, as the PA signal due to polaron absorption is strongly fluence dependent due to two-step exciton dissociation by resonant sequential excitation.\cite{Silva01}

\begin{figure}
\includegraphics{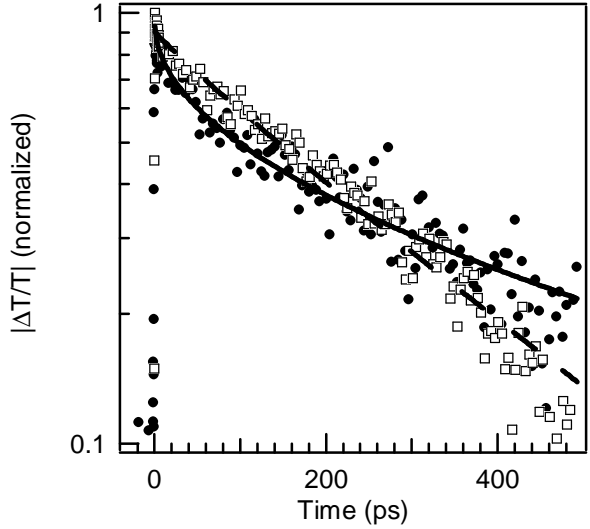}
\caption{\label{fig:PalowP} Absorption transients at a probe photon energy of 1.46\,eV  and solution temperature of 14$^{\circ}$C (21\,$\mu$J\,cm$^{-2}$ pump fluence, black circles) and 65$^{\circ}$C (25\,$\mu$J\,cm$^{-2}$ pump fluence, open squares) with fits to the expression ($c \exp(-\frac{t}{\tau}-(\frac{t}{a})^b)$. The solid line is fit to the data at 14$^{\circ}$C ($a=300$\,ps, $b=0.4$ and $\tau$=1600\,ps) and the broken line is the fit to the data at 65$^{\circ}$C ($a=300$\,ps, $b=0.9$ and $\tau$=1600\,ps).}
\end{figure}

Figure~\ref{fig:PalowP} shows the time evolution of transient PA signal recorded at a probe photon energy of 1.46\,eV for low excitation fluence ($\sim 20$\,$\mu$J\,cm$^{-2}$). This corresponds to the regime where the PL intensity depends linearly on the excitation fluence (see the arrows in Figure \ref{fig:PLPow}) and we can therefore neglect exciton-exciton interactions. The transient signal measured at 65$^{\circ}$C (open squares) decays quasi-exponentially while the decay of the 14$^{\circ}$C signal (filled circles) displays a much faster initial component. The exponential component is assigned to the decay of localized excitons on isolated oligomers. We have shown previously that well below the transition temperature, exciton dynamics are dominated by fast diffusion within the chiral supramolecular assembly, which leads to trapping and subsequent slower transfer of localized excitons.\cite{Herz03a} The ultrafast decay of the signal measured at 14$^{\circ}$C can therefore be interpreted as a consequence of exciton diffusion along the stacks.

\begin{figure}
\includegraphics{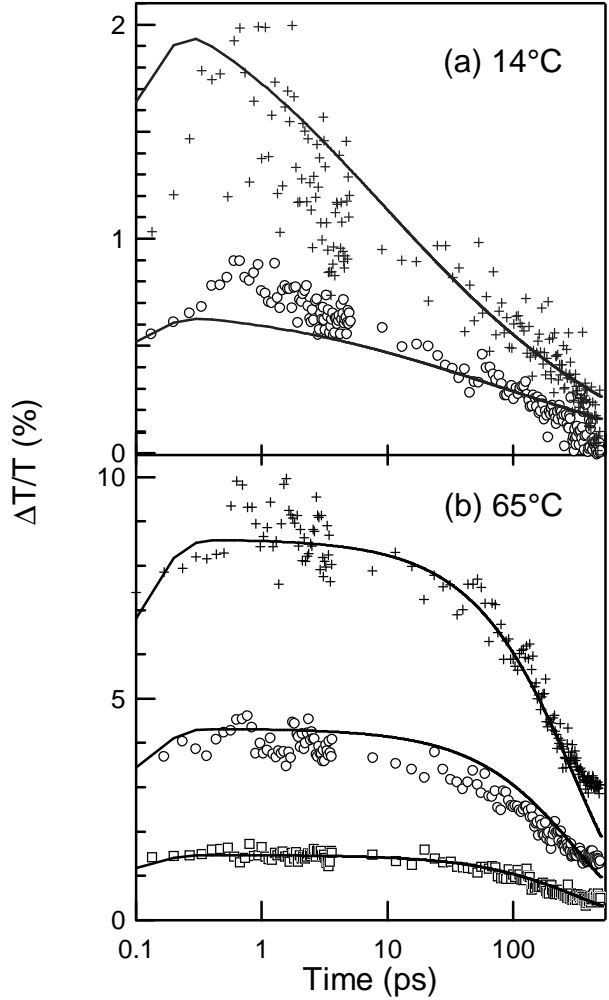}
\caption{\label{fig:SE} (a) SE signal (probe energy 2.21\,eV) at
14$^{\circ}$C for 279\,$\mu$J\,cm$^{-2}$ (crosses) and
111\,$\mu$J\,cm$^{-2}$ (circles) pump fluence. The lines are fits with the photophysical model at the corresponding pump fluence. (b) SE signal (probe
energy 2.21\,eV) at 65$^{\circ}$C at 836\,$\mu$J\,cm$^{-2}$
(crosses), 209\,$\mu$J\,cm$^{-2}$ (circles) and 70\,$\mu$J\,cm$^{-2}$
(squares) pump fluence. The lines are fits to the model at the
corresponding pump fluence.}
\end{figure}

\begin{figure}
\includegraphics{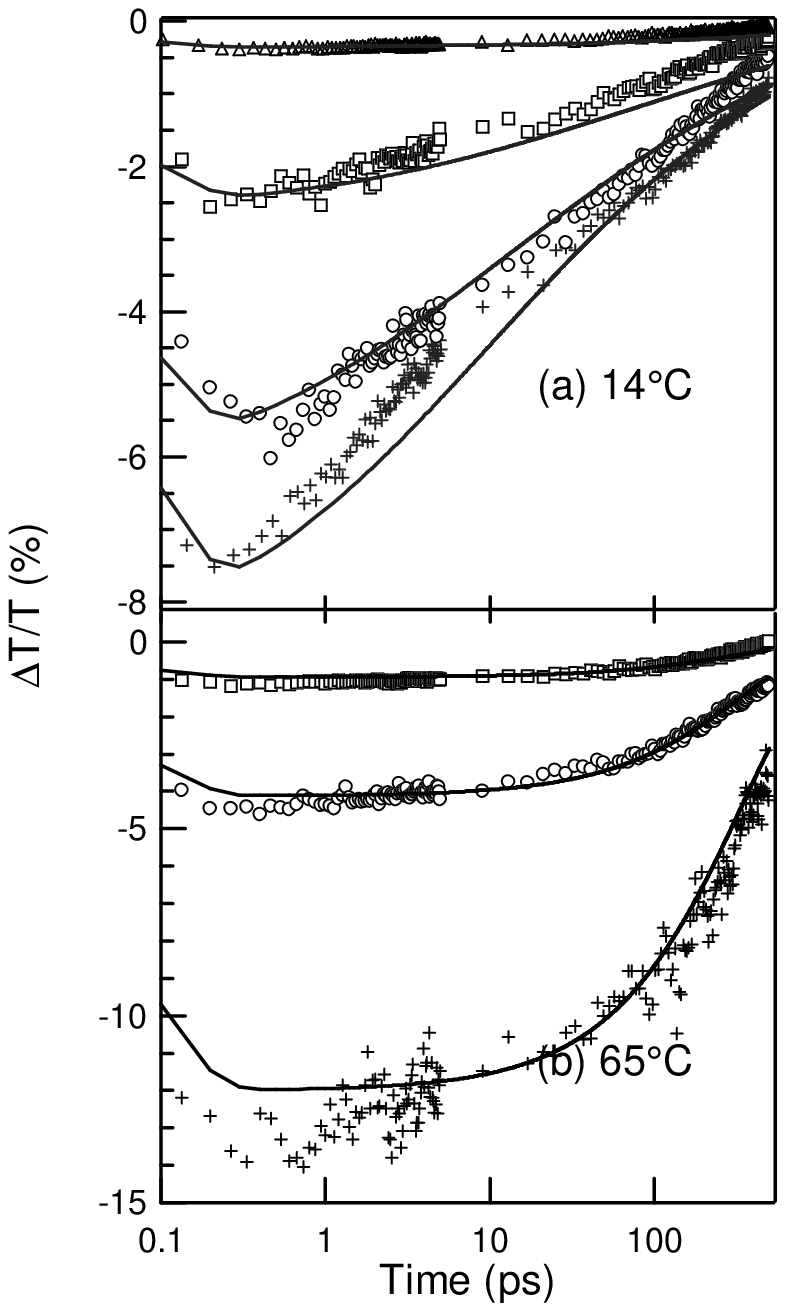}
\caption{\label{fig:PIA} (a) PA signals (probe energy 1.46\,eV) at
14$^{\circ}$C for pump fluence of 557\,$\mu$J\,cm$^{-2}$ (crosses),
209\,$\mu$J\,cm$^{-2}$ (circles), 84\,$\mu$J\,cm$^{-2}$ (squares) and
21\,$\mu$J\,cm$^{-2}$ (triangles). (b) PA signals (probe energy
1.46\,eV) at 65$^{\circ}$C for pump fluence of
557\,$\mu$J\,cm$^{-2}$ (crosses), 111\,$\mu$J\,cm$^{-2}$ (circles) and
25\,$\mu$J\,cm$^{-2}$ (squares). The lines are the fits to the data
at corresponding pump fluence according to the model described in
the text.}
\end{figure}

\begin{figure}
\includegraphics{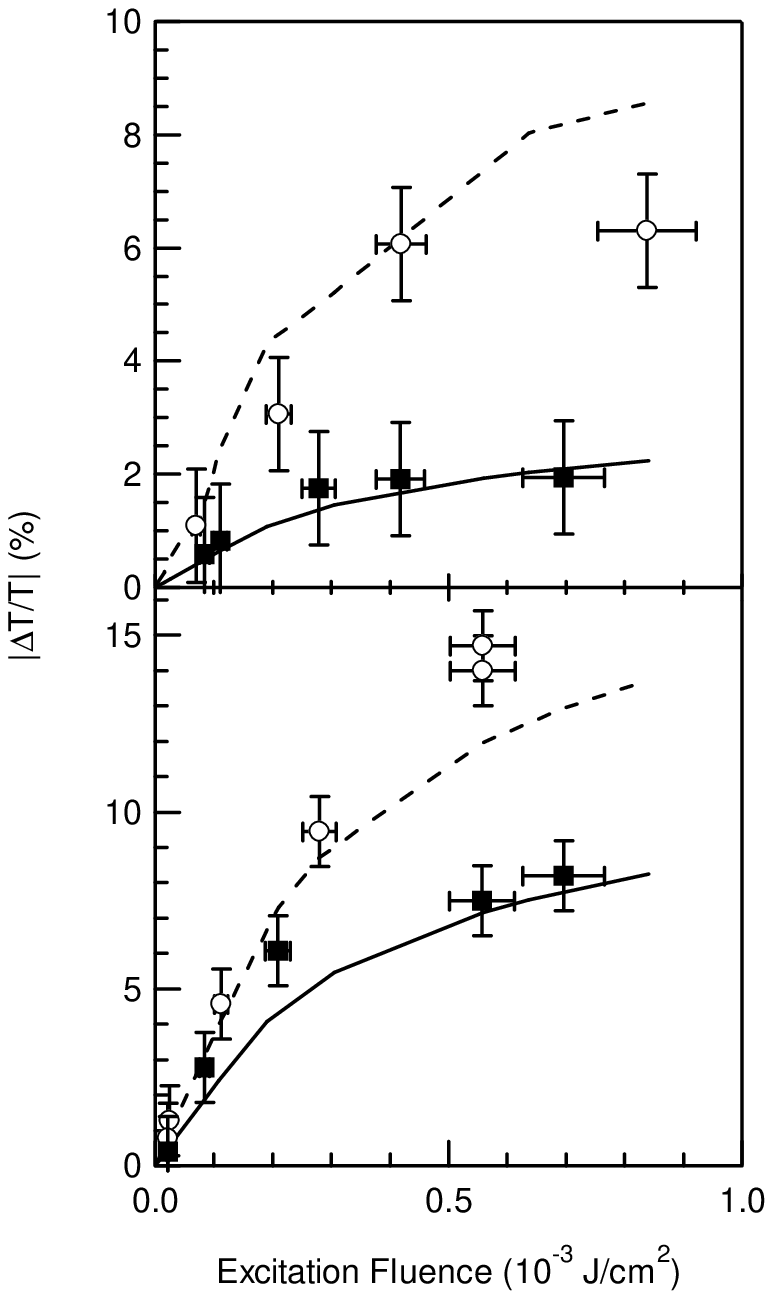}
\caption{\label{fig:PASE0} (a) SE signal (probe energy 2.21\,eV) at $t = 0$ as a function of excitation fluence at 14$^{\circ}$C (black squares) and at 65$^{\circ}$C (open circles). This amplitude was determined by fitting the transient absorption data in Figures~\ref{fig:SE} and~\ref{fig:PIA} with a convolution of the instrument response function and a multiexponential function, and extrapolating to zero time. This procedure allows accurate deconvolution of the instrument response. (b) PA signals (probe energy 1.46\,eV) at $t = 0$ as a function of excitation fluence at 14$^{\circ}$C (black squares) and at 65$^{\circ}$C (open circles), determined as described above. The lines through all data are fits to the photophysical model described in the text.}
\end{figure}

We have measured time-dependent SE and PA transient signal in MOPV4 solution at various pump fluences, probing at 2.21\,eV (Figure~\ref{fig:SE}) and 1.46\,eV (Figure~\ref{fig:PIA}), respectively. The dynamics of both signals measured at 14$^{\circ}$C depend on the excitation fluence, and at higher fluence are considerably faster than those measured at 65$^{\circ}$C. The fast decay present within the first 10\,ps at 14$^{\circ}$C increases in relative contribution and becomes faster as the excitation fluence is increased. At those fluences, the initial PA and SE signals (at $t = 0$) have a sub-linear dependence in the excitation fluence (Figure~\ref{fig:PASE0}). The deviation from linearity is stronger in the data measured at 14$^{\circ}$C than for those measured at 65$^{\circ}$C. Also, the transition between linear and sub-linear regimes at 14$^{\circ}$C occurs at lower fluence. These results are consistent with the fluence dependence of the integrated PL intensity (Figure~\ref{fig:PLPow}).

The sub-linear excitation fluence dependence displayed in Figures~\ref{fig:PLPow} and~\ref{fig:PASE0} are characteristic of bimolecular exciton annihilation dynamics. Since the transient absorption signal decay rates depend on the excitation fluence at 14$^{\circ}$C, we correlate the exciton-exciton annihilation processes with the formation of supramolecular assemblies. At 14$^{\circ}$C, exciton diffusion on the supramolecular stacks is fast enough to allow for exciton-exciton interactions at sufficiently high density. At 65$^{\circ}$C, localization of excitons becomes dominant as the disorder is increased.\cite{Herz03a} Furthermore, just below the transition temperature, the average stack length is shorter as the equilibrium shifts towards the dissolved phase\cite{Jonkheijm03} so that the exciton density required for exciton-exciton interaction is higher.

In order to obtain a more quantitative description of exciton bimolecular annihilation dynamics in MOPV4 nanostructures, we have constructed a mesoscopic model that reproduces our experimental results and allows us to unravel the mechanism governing the annihilation process. The model is based on a rate equation for the exciton population, $N$, given by

\begin{eqnarray}
\frac{dN}{dt} = G(t) - \frac{N}{\tau} - \beta N^2 - \gamma \frac{N^2}{t^d} - \epsilon N \left(\frac{t}{a}\right)^{b-1}
\label{eq:model}
\end{eqnarray}
The first term is the exciton generation rate, which has the form

\begin{eqnarray}
G(t)=\frac{\alpha P \lambda}{2 \pi r^2 h c} \frac{(N_{gr}-S(t))}{N_{gr}} \frac{\exp(-(t-t_0)^2/2 \sigma_t^2)}{\sqrt{2 \pi \sigma_t^2}}
\end{eqnarray}
where $P$ is the excitation pulse energy, $\alpha$ is the
unsaturated absorbtion coefficient, $\lambda$ is the excitation
wavelength, $r$ is the Gaussian radius of the excitation beam,
$S(t)$ is the number of excitons created between $t_0$ and $t$
(with $t_0$ the zero-time), $N_{gr}$ is the initial density of
ground states and $\sigma_t$ is the width of the excitation pulse.
To calculate $S(t)$, we use
\begin{eqnarray}
S(t+dt)= G(t) dt
\end{eqnarray}
The generation term takes into account the Gaussian distribution in space and time of the excitation pulse. The initial density of ground states is taken to be the concentration of MOPV4 molecules with the ground state depletion taken into account in each point.

The second term in equation~\ref{eq:model} describes the unimolecular excited-state decay rate, with time constant $\tau$ measured using TCSPC for an MOPV4 solution at 14 and 65$^{\circ}$C at a very low excitation fluence. Collisional bimolecular annihilation processes are described by the third term, where $\beta$ is the diffusion-limited exciton bimolecular annihilation rate constant. The general form includes a time-dependence but this term may be neglected if the typical diffusion length on experimental time-scales is larger than the extent of the exciton.\cite{Powell75,Dogariu98} In conjugated systems, this is usually the case; however, this assumption will be discussed further in Section~\ref{discu}.

On the other hand, the long-range time-dependent bimolecular annihilation term, with rate constant $\gamma$, includes explicit time-dependence. This time-dependence was introduced following resonance energy transfer studies and accounts for the dispersive nature of the transfer process resulting from the distribution of transfer rates in a random ensemble of donors and acceptors, which leads to a non-Markovian depletion of acceptors near the donors. The exact time dependence, characterized by the time exponent $d$, varies with the spatial distribution of acceptors and depends on dimensionality.\cite{Forster49,Eisenthal64} However, other factors such as disorder may also lead to a power-law dependence of the transfer rate (this point will be discussed further in Section~\ref{discu}) and we have therefore considered $d$ in this term as a fitting variable.

The final term in equation~\ref{eq:model} represents the ultrafast dynamics related to exciton diffusion in supramolecular assemblies. We have reported previously dispersive diffusion dynamics of excitons in MOPV4 and we will therefore not consider these issues in detail here.\cite{Herz03a} To take into account effects of diffusion to non-radiative traps, we have added a term in our rate equation so that at low excitation fluence, the solution converges towards the low excitation fluence data. To determine this term, we fit the PA signal measured at the lowest excitation fluence (within the linear intensity regime, see Figure~\ref{fig:PalowP}) with a stretched exponential taking into account the natural decay ($\exp(-t/\tau-(t/a)^b)$), where $a$ is the stretched-exponential time constant and $b$ is the exponent in the stretched exponential. In equation~\ref{eq:model}, $\epsilon$ is the amplitude of the contribution of this dispersive process. This method follows theoretical\cite{Balagurov74,Movaghar86,Mollay94} and experimental\cite{Yan94,Herz00} studies of exciton diffusion to traps were a stretched exponential behavior was obtained. The stretched exponential fit to the 65$^{\circ}$C PA data is very close to an exponential function but the time dependence of the PA data at 14$^{\circ}$C is close to an exponential of the square root of time. This indicates that diffusion to traps and polarization decay influence the PA signal dynamics at 14$^{\circ}$C but is less important at 65$^{\circ}$C, as established by our previous studies of exciton dynamics on MOPV4 at low fluence.\cite{Herz03a}

The experimental geometry is taken into account in the photophysical model by dividing the volume located at the intersection of the probe and the pump beams into small sub-volumes with homogeneous excitation fluence where the rate equation can be solved. The measured quantities are then calculated by summing the exciton densities in a mesoscopic approach. To calculate the SE and PA signals, the fraction of probe photons transmitted through the sample needs to be evaluated. This fraction is related to the exciton density and to the SE and PA cross-sections by
\begin{eqnarray}
d \rho_{\nu} = \pm \sigma N \rho_{\nu} dx
\end{eqnarray}
where $\rho_{\nu}$ is the photon density, $\sigma$ is the cross-section for SE or for PA and $x$ is the sample depth. As we are only interested in the sample transmission, the initial probe photon population is arbitrary. The average exciton density at a particular sample depth is determined in a similar manner as the total exciton density on the entire volume but the multiplication factors needed are the surface fractions excited with each of the particular excitation fluence of the set, at the particular sample depth.

The parameter $\tau$ is obtained separately from TCSPC measurements, $N_{gr}$ from the knowledge of the MOPV4 concentration, $\alpha$ and therefore $\sigma$ (defined as $\alpha/N_{gr}$) from UV-vis absorption measurements, and $a$ and $b$ from the stretched-exponential fitting of the low fluence PA signal. To determine the remaining parameters ($\beta$, $\gamma$ and $d$, $\epsilon$, $\sigma_{PA}$ and $\sigma_{SE}$), the PA and SE signals at all excitation fluence, the zero-time signals and the integrated PL data are modelled using a global fit. The best results of the fitting procedure are displayed in Figures~\ref{fig:PLPow}, \ref{fig:SE}, \ref{fig:PIA} and \ref{fig:PASE0} with the model parameters summarized in Table~\ref{tab:Param}.

\begin{figure}
\includegraphics{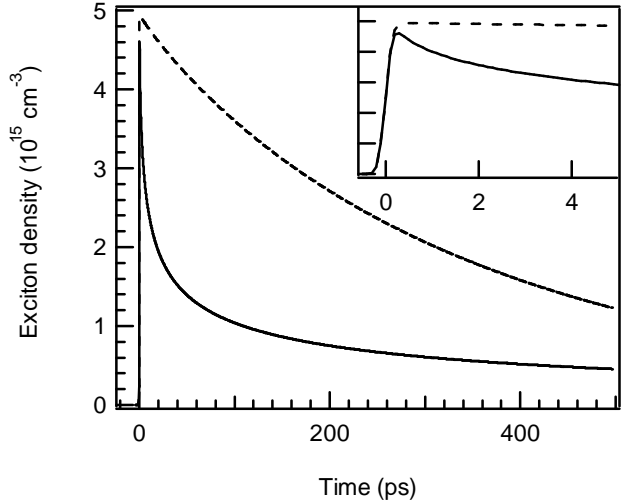}
\caption{\label{fig:ExcM} Exciton density at peak excitation fluence
against time for the parameters reported in Table~\ref{tab:Param} at
14$^{\circ}$C (solid line) and 65$^{\circ}$C (broken line). Inset:
Data plotted over the first 5\,ps.}
\end{figure}

\begin{table}
\caption{\label{tab:Param} Summary of parameters used in the
photophysical model.}
\begin{ruledtabular}
\begin{tabular}{c c c}
& 14$^{\circ}$C & 65$^{\circ}$C \\
\hline
 $\tau$ (ps) & 2940 & 2940 \\
 $\beta$ (cm$^3$\,ps$^{-1}$)& 0 & 0 \\
 $\gamma$ (cm$^3$\,ps$^{d-1}$)& 3.6 $\times 10^{-17}$ & 0 \\
 $d$ & 0.5 & - \\
 $\epsilon$ (ps$^{-1}$) & 2.2 $\times 10^{-4}$ & 2.5 $\times 10^{-3}$\\
 $a$ (ps) & 266 & 300 \\
 $b$ & 0.46 & 0.9 \\
 $\sigma$ (cm$^2$) & 9.25 $\times 10^{-17}$ &  9.25 $\times 10^{-17}$\\
 $N_{gr}$ (cm$^{-3}$) & 2.49 $\times 10^{17}$ &  2.49 $\times 10^{17}$\\
 $\sigma_{PA}$ (cm$^2$) & 1.55 $\times 10^{-16}$ & 2.5 $\times 10^{-16}$\\
 $\sigma_{SE}$ (cm$^2$) & 4 $\times 10^{-17}$ &  1.4 $\times 10^{-16}$\\
\end{tabular}
\end{ruledtabular}
\end{table}
A high bimolecular annihilation rate constant $\gamma$ was needed to fit the time-resolved data taken at 14$^{\circ}$C, but bimolecular annihilation effects were not necessary to reproduce the data taken at 65$^{\circ}$C. We find that the magnitude of $\gamma$ is almost an order of magnitude higher than the typical values obtained in polymer films.\cite{Dogariu98,Vacar97,Maniloff97,Cerullo98} Furthermore, the data at 14$^{\circ}$C cannot be fitted with a time-independent annihilation term, and setting $\beta = 0$ did not compromise the global fits presented here. Variation of the exponent $d$ between 0.4 and 0.7 does not have a significant effect on the quality of the fit; however, the best fit is obtained with a square root dependence in time ($d=0.5$). Figure~\ref{fig:ExcM} shows the exciton density at the maximum experimental fluence used here ($\sim 840$\,$\mu$J\,cm$^{-2}$), as determined by the model. The time-dependent bimolecular annihilation term leads to an ultrafast decay of the exciton density at 14$^{\circ}$C, but is insignificant at 65$^{\circ}$C. Note that the peak exciton population density at both temperatures ($<5 \times 10^{15}$\,cm$^{-3}$) is 1.8\% of the ground-state chromophore density used in these experiments ($N_{gr}$ in Table~\ref{tab:Param}), and therefore photophysical saturation is not the origin of sublinear fluence dependence observed in Figures~\ref{fig:PLPow} and \ref{fig:PASE0}. Such low exciton densities in the supramolecular nanostructure point to the need for a high bimolecular annihilation rate constant at the photon flux regime used here. 

\section{Discussion \label{discu}}
Our model shows that exciton bimolecular annihilation is necessary to describe excited-state dynamics in MOPV4 solution at 14$^{\circ}$C while it has a negligible effect at 65$^{\circ}$C. To check that the high value we obtain for the bimolecular annihilation constant is not a consequence of the mesoscopic character of our model, we performed a similar analysis to those developed in a range of previous studies.\cite{Vacar97,Maniloff97,Shimizu01} In this crude analysis, the exciton density was modelled with a simplified rate equation with a unimolecular decay term, a time-dependent bimolecular annihilation term and a delta function as the generation term.
\begin{eqnarray}
\frac{d N}{d t} & = & N_0\,\delta(t) -\frac{N}{\tau} - \gamma \frac{N^2}{\sqrt{t}}\\
N & = & \frac{\exp\left(-t/\tau\right)}{\left(
N_{0}^{-1}+2 \gamma \sqrt{\tau}
\int_0^{\sqrt{t/\tau}}\exp(-x^2)dx \right)}
\end{eqnarray}
For small signals, the PA signal is directly related to the exciton population ($\Delta T/T \approx -\sigma_{PA} Nx$), so we also fitted our data with the analytical solution of the simplified rate equation. The exciton population at zero-time, $N_0$, was approximated with the known density of absorbed photons, which allowed to relate the fitting coefficient to the parameter $\gamma$. The value obtained from the high fluence PA signals, $\gamma \approx 2 \times 10^{-17}$\,cm$^3$\,ps$^{-1/2}$, is very similar to the result of our model, which is reproduced by this crude analysis but is closer to the MOPV4 physics as it takes into account the variation in excitation fluence in the sample.

Several studies have interpreted the square-root-of-time dependence of the bimolecular annihilation as an effect of the non-Markovian depletion of close acceptor-donor pairs. With a point-dipole approximation, the F\"{o}rster model predicts a time dependence of the excitation transfer rate of $t^{-(1-D/6)}$, with $D$ being the dimensionality of the acceptor distribution. This result was obtained by applying the method developed by Eisenthal~\etal~for the three-dimensional case to the one- and two-dimensional cases.\cite{Eisenthal64} If this model is valid to describe the exciton-exciton annihilation, the dimensionality characteristic of exciton distribution in MOPV4 stacks is three (which corresponds to the best fit, with $d=0.5$, but a fractal dimensionality between three and two cannot be ruled out). The point-dipole model also relates the magnitude of exciton-exciton annihilation rate constant to the F\"{o}rster radius by
\begin{eqnarray}
\gamma^D &=& \frac{R_0^D\, \pi^{D/2} \,
\Gamma\left(2-D/6\right)}{\Gamma\left(1+D/2\right)\, \xi^{D/6}}\\
\gamma^3 &=& \frac{2\,\pi}{3}\,R_0^3\,\sqrt{\frac{\pi}{\xi}}
\end{eqnarray}
$\Gamma$ is the Gamma function, $\xi$ the exciton lifetime in the absence of transfer and $R_0$ the F\"{o}rster radius. With $D=3$, we extract $R_0 \approx$ 70\,nm with our value of $\gamma$, which is unrealistic. However, the simple model is based on a random distribution of acceptors in a sphere of dimensionality $D$ centered on the donor and it fails to reproduce a distribution of excitons in the complex geometry of MOPV4 stacks. We also point out the limitations of the applicability of F\"{o}rster theory in organic semiconductor systems, where the point dipole approximation is often unjustified.\cite{Beljonne02}

The collisional mechanism used in the model does not include any time-dependence. This is justified if the typical diffusion length on experimental time-scales is larger than the spatial extent of the exciton. However, the general expression in three dimensions includes a time-dependence given by\cite{Powell75}
\begin{eqnarray}
\beta = 8 \pi R_e D \left(1+\frac{R_e}{\sqrt{\pi D t}}\right)
\end{eqnarray}
with $D$ the diffusion coefficient and $R_e$ the collision distance. If we assign the observed square-root time dependence to this simple collisional mechanism, we obtain an unrealistic value for the diffusion coefficient $D \approx 10^5$\,cm$^2$\,s$^{-1}$ with $R_e = 1$\,nm. The collision radius $R_e$ would have to be around 50\,nm for $D$ to be similar to the values found in organic systems ($D \approx 10^{-2}$\,cm$^2$\,s$^{-1}$).\cite{Powell75} Furthermore, such high values of $D$ correspond to the time-independent regime so that the collisional mechanism cannot explain our data. If we use the value of $\beta$ corresponding to the best global fit (which is significantly worse than that shown in Figures~\ref{fig:PLPow}, \ref{fig:SE}, \ref{fig:PIA}, and \ref{fig:PASE0}), realistic values for the diffusion coefficient can be obtained if $R_e \geq 5$\,nm.\cite{Powell75} The expression used here corresponds to diffusion in a three-dimensional lattice, but similar expressions exist for the one and two-dimensions cases.\cite{Suna70} All of these fail to reproduce the observed bimolecular exciton annihilation time-dependence and intensity dependence.

With the above considerations, we can conclude that neither the collisional model nor the F\"{o}rster (point dipole) model can uniquely explain the high rate of exciton bimolecular annihilation measured in MOPV4 stacks. These two mechanisms represent limiting cases where either incoherent exciton hopping or F\"{o}rster energy transfer are sufficient to describe the annihilation. A more realistic description is an intermediate case of fast incoherent hopping followed by efficient resonance energy transfer. The square-root-of-time dependence of the bimolecular annihilation term extracted from the model in equation~\ref{eq:model} is an effect of the complex geometry of the supramolecular assemblies and of disorder. Combined with rapid exciton diffusion within the stack,\cite{Herz03a} bimolecular interactions lead to the high annihilation rate. A branching between exciton localization and direct bimolecular annihilation is likely at early time. At later times, bimolecular annihilation of localized excitons is likely. This picture is analogous to related studies of resonance energy transfer dynamics in blends containing small concentrations of MOPV4 in similar oligomers with shorter conjugation length.\cite{Hoeben03} 

A more accurate description of the bimolecular annihilation dynamics would require comparing the time-dependent average exciton density accessed with our measurement and a configurationally averaged microscopic description of the transfer processes using a Monte-Carlo or a master equation approach. Implementation of these schemes requires a significant number of assumptions, including detailed understanding of the structural and geometrical disorder. This information is beyond the scope of the present study.

On the other hand, a detailed description should be based on a quantum-chemical, non-Markovian model where the exciton-exciton coupling and electron-phonon coupling are explicitly taken into account. OPV systems have been shown to belong to the intermediate coupling regime where the relaxation energy accompanying optical excitation, the nearest neighbor exciton-exciton coupling, and the frequency of the most strongly coupled intramolecular vibration are comparable.\cite{Spano02} In MOPV4 supramolecular stacks, the intermolecular exciton-exciton coupling are comparable to typical intramolecular exciton-exciton coupling found in polymeric systems.\cite{BeljonneC} For such an intermediate case between polymeric systems and molecular aggregates that undergo coherent excitation transfer (supertransfer), the description of exciton-exciton interaction would need to go beyond a F\"{o}rster description of energy transfer.\cite{Scholes01,Scholes02}

We are currently exploring both the microscopic semi-classical and the quantum chemical approaches. However, the work presented here demonstrates unambiguously that the exciton dynamics in supramolecular nanostructures in solution are similar to those observed in polymeric semiconductors, but control of intermolecular electronic interactions is higher such that high exciton bimolecular annihilation rates are measured.

\section{Conclusion}
We have shown that exciton dynamics in supramolecular assemblies of MOPV4 are very similar to the dynamics observed in conjugated polymer films. Supramolecular chemistry is a very promising route to construct well-defined semiconductor architectures with polymeric properties through controlled reversible assembly.

We have applied femtosecond transient absorption spectroscopy as a means to explore exciton dynamics in a supramolecular nanostructure constructed with a hydrogen-bonded oligophenylenevinylene derivative. A simple rate-equation model was applied successfully to separate exciton bimolecular annihilation phenomena from unimolecular processes occurring at low fluence.\cite{Herz03a} We have shown that exciton-exciton annihilation does take place on the supramolecular assemblies and dominates the exciton dynamics at high fluence. The mechanism for bimolecular annihilation is non-Markovian; explicit time dependence is necessary to reproduce the experimental data with the model. The rate constant for bimolecular annihilation is higher than in comparable organic systems and cannot be rationalized with simple models used in previous studies. We consider that the electronic properties of this system fall in the intermediate electronic coupling regime, where intermolecular electronic coupling is comparable to the intramolecular reorganization energy, providing an opportunity to apply semiclassical and quantum-chemical methodologies to explore electronic dynamics in this promising regime of supramolecular electronics. We are currently pursuing this opportunity.


\begin{acknowledgments}
The authors thank Prof. Gregory Scholes, Dr. Stefan Meskers, and Dr. David Beljonne for fruitful discussions. The work in Cambridge is supported by the UK Engineering and Physical Sciences Research Council (EPSRC) and the interdisciplinary research center for nanotechnology (Cambridge, UCL, Bristol).  CD is an Isaac Newton Scholar. CS is an EPSRC Advanced Research Fellow. LMH thanks St John's College, Cambridge for financial support. The Work in Eindhoven is supported by the Netherlands Organization for Scientific Research (NWO, CW). APHJS is supported with a Fellowship of the Royal Netherlands Academy of Arts and Science. The Cambridge-Eindhoven collaboration is supported by the European Commission (LAMINATE).
\end{acknowledgments}



\begin{thebibliography}{51}
\expandafter\ifx\csname natexlab\endcsname\relax\def\natexlab#1{#1}\fi
\expandafter\ifx\csname bibnamefont\endcsname\relax
  \def\bibnamefont#1{#1}\fi
\expandafter\ifx\csname bibfnamefont\endcsname\relax
  \def\bibfnamefont#1{#1}\fi
\expandafter\ifx\csname citenamefont\endcsname\relax
  \def\citenamefont#1{#1}\fi
\expandafter\ifx\csname url\endcsname\relax
  \def\url#1{\texttt{#1}}\fi
\expandafter\ifx\csname urlprefix\endcsname\relax\def\urlprefix{URL }\fi
\providecommand{\bibinfo}[2]{#2}
\providecommand{\eprint}[2][]{\url{#2}}

\bibitem[{\citenamefont{Stutzmann et~al.}(2003)\citenamefont{Stutzmann, Friend,
  and Sirringhaus}}]{Stutzmann03}
\bibinfo{author}{\bibfnamefont{N.}~\bibnamefont{Stutzmann}},
  \bibinfo{author}{\bibfnamefont{R.~H.} \bibnamefont{Friend}},
  \bibnamefont{and}
  \bibinfo{author}{\bibfnamefont{H.}~\bibnamefont{Sirringhaus}},
  \bibinfo{journal}{Science} \textbf{\bibinfo{volume}{299}},
  \bibinfo{pages}{1881} (\bibinfo{year}{2003}).

\bibitem[{\citenamefont{Dimitrakopoulos and
  Malenfant}(2002)}]{Dimitrakopoulos02}
\bibinfo{author}{\bibfnamefont{C.~D.} \bibnamefont{Dimitrakopoulos}}
  \bibnamefont{and} \bibinfo{author}{\bibfnamefont{P.~R.~L.}
  \bibnamefont{Malenfant}}, \bibinfo{journal}{Adv. Mater.}
  \textbf{\bibinfo{volume}{14}}, \bibinfo{pages}{99} (\bibinfo{year}{2002}).

\bibitem[{\citenamefont{Friend et~al.}(1999)\citenamefont{Friend, Gymer,
  Holmes, Burroughes, Marks, Taliani, Bradley, Santos, Br\'{e}das, L\"{o}gdlund
  et~al.}}]{Friend99}
\bibinfo{author}{\bibfnamefont{R.~H.} \bibnamefont{Friend}},
  \bibinfo{author}{\bibfnamefont{R.~W.} \bibnamefont{Gymer}},
  \bibinfo{author}{\bibfnamefont{A.~B.} \bibnamefont{Holmes}},
  \bibinfo{author}{\bibfnamefont{J.~H.} \bibnamefont{Burroughes}},
  \bibinfo{author}{\bibfnamefont{R.~N.} \bibnamefont{Marks}},
  \bibinfo{author}{\bibfnamefont{C.}~\bibnamefont{Taliani}},
  \bibinfo{author}{\bibfnamefont{D.~D.~C.} \bibnamefont{Bradley}},
  \bibinfo{author}{\bibfnamefont{D.~A.~D.} \bibnamefont{Santos}},
  \bibinfo{author}{\bibfnamefont{J.~L.} \bibnamefont{Br\'{e}das}},
  \bibinfo{author}{\bibfnamefont{M.}~\bibnamefont{L\"{o}gdlund}},
  \bibnamefont{et~al.}, \bibinfo{journal}{Nature}
  \textbf{\bibinfo{volume}{397}}, \bibinfo{pages}{121} (\bibinfo{year}{1999}).

\bibitem[{\citenamefont{Forrest et~al.}(2000)\citenamefont{Forrest, Burrows,
  and Thompson}}]{Forrest00}
\bibinfo{author}{\bibfnamefont{S.}~\bibnamefont{Forrest}},
  \bibinfo{author}{\bibfnamefont{P.}~\bibnamefont{Burrows}}, \bibnamefont{and}
  \bibinfo{author}{\bibfnamefont{M.}~\bibnamefont{Thompson}},
  \bibinfo{journal}{IEEE Spectrum} \textbf{\bibinfo{volume}{37}},
  \bibinfo{pages}{29} (\bibinfo{year}{2000}).

\bibitem[{\citenamefont{Brabec et~al.}(2002)\citenamefont{Brabec, Winder,
  Sariciftci, Hummelen, Dhanabalan, van Hal, and Janssen}}]{Brabec02}
\bibinfo{author}{\bibfnamefont{C.~J.} \bibnamefont{Brabec}},
  \bibinfo{author}{\bibfnamefont{C.}~\bibnamefont{Winder}},
  \bibinfo{author}{\bibfnamefont{N.~S.} \bibnamefont{Sariciftci}},
  \bibinfo{author}{\bibfnamefont{J.~C.} \bibnamefont{Hummelen}},
  \bibinfo{author}{\bibfnamefont{A.}~\bibnamefont{Dhanabalan}},
  \bibinfo{author}{\bibfnamefont{P.~A.} \bibnamefont{van Hal}},
  \bibnamefont{and} \bibinfo{author}{\bibfnamefont{R.~A.~J.}
  \bibnamefont{Janssen}}, \bibinfo{journal}{Adv. Funct. Mater.}
  \textbf{\bibinfo{volume}{12}}, \bibinfo{pages}{709} (\bibinfo{year}{2002}).

\bibitem[{\citenamefont{Gregg}(2003)}]{Gregg03}
\bibinfo{author}{\bibfnamefont{B.~A.} \bibnamefont{Gregg}},
  \bibinfo{journal}{J. Phys. Chem. B} \textbf{\bibinfo{volume}{107}},
  \bibinfo{pages}{4688} (\bibinfo{year}{2003}).

\bibitem[{\citenamefont{Peumans et~al.}(2003)\citenamefont{Peumans, Yakimov,
  and Forrest}}]{Peumans03}
\bibinfo{author}{\bibfnamefont{P.}~\bibnamefont{Peumans}},
  \bibinfo{author}{\bibfnamefont{A.}~\bibnamefont{Yakimov}}, \bibnamefont{and}
  \bibinfo{author}{\bibfnamefont{S.~R.} \bibnamefont{Forrest}},
  \bibinfo{journal}{J. Appl. Phys.} \textbf{\bibinfo{volume}{93}},
  \bibinfo{pages}{3693} (\bibinfo{year}{2003}).

\bibitem[{\citenamefont{Setayesh et~al.}(2001)\citenamefont{Setayesh,
  Grimsdale, Weil, Enkelmann, M\"{u}llen, Meghdadi, List, and
  Leising}}]{Setayesh01}
\bibinfo{author}{\bibfnamefont{S.}~\bibnamefont{Setayesh}},
  \bibinfo{author}{\bibfnamefont{A.~C.} \bibnamefont{Grimsdale}},
  \bibinfo{author}{\bibfnamefont{T.}~\bibnamefont{Weil}},
  \bibinfo{author}{\bibfnamefont{V.}~\bibnamefont{Enkelmann}},
  \bibinfo{author}{\bibfnamefont{K.}~\bibnamefont{M\"{u}llen}},
  \bibinfo{author}{\bibfnamefont{F.}~\bibnamefont{Meghdadi}},
  \bibinfo{author}{\bibfnamefont{E.~J.~W.} \bibnamefont{List}},
  \bibnamefont{and} \bibinfo{author}{\bibfnamefont{G.}~\bibnamefont{Leising}},
  \bibinfo{journal}{J. Am. Chem. Soc.} \textbf{\bibinfo{volume}{123}},
  \bibinfo{pages}{946} (\bibinfo{year}{2001}).

\bibitem[{\citenamefont{Ego et~al.}(2003)\citenamefont{Ego, Marsitzky, Becker,
  Zhang, Grimsdale, M\"{u}llen, Mackenzie, Silva, and Friend}}]{Ego03}
\bibinfo{author}{\bibfnamefont{C.}~\bibnamefont{Ego}},
  \bibinfo{author}{\bibfnamefont{D.}~\bibnamefont{Marsitzky}},
  \bibinfo{author}{\bibfnamefont{S.}~\bibnamefont{Becker}},
  \bibinfo{author}{\bibfnamefont{J.~Y.} \bibnamefont{Zhang}},
  \bibinfo{author}{\bibfnamefont{A.~C.} \bibnamefont{Grimsdale}},
  \bibinfo{author}{\bibfnamefont{K.}~\bibnamefont{M\"{u}llen}},
  \bibinfo{author}{\bibfnamefont{J.~D.} \bibnamefont{Mackenzie}},
  \bibinfo{author}{\bibfnamefont{C.}~\bibnamefont{Silva}}, \bibnamefont{and}
  \bibinfo{author}{\bibfnamefont{R.~H.} \bibnamefont{Friend}},
  \bibinfo{journal}{J. Am. Chem. Soc.} \textbf{\bibinfo{volume}{125}},
  \bibinfo{pages}{437} (\bibinfo{year}{2003}).

\bibitem[{\citenamefont{Grimsdale et~al.}(2002)\citenamefont{Grimsdale,
  Lecl\`{e}re, Mackenzie, Murphy, Setayesh, Silva, Friend, and
  M\"{u}llen}}]{Grimsdale02}
\bibinfo{author}{\bibfnamefont{A.~C.} \bibnamefont{Grimsdale}},
  \bibinfo{author}{\bibfnamefont{P.}~\bibnamefont{Lecl\`{e}re}},
  \bibinfo{author}{\bibfnamefont{J.~D.} \bibnamefont{Mackenzie}},
  \bibinfo{author}{\bibfnamefont{C.}~\bibnamefont{Murphy}},
  \bibinfo{author}{\bibfnamefont{S.}~\bibnamefont{Setayesh}},
  \bibinfo{author}{\bibfnamefont{C.}~\bibnamefont{Silva}},
  \bibinfo{author}{\bibfnamefont{R.~H.} \bibnamefont{Friend}},
  \bibnamefont{and}
  \bibinfo{author}{\bibfnamefont{K.}~\bibnamefont{M\"{u}llen}},
  \bibinfo{journal}{Adv. Funct. Mater.} \textbf{\bibinfo{volume}{12}},
  \bibinfo{pages}{729} (\bibinfo{year}{2002}).

\bibitem[{\citenamefont{Sirringhaus et~al.}(2000)\citenamefont{Sirringhaus,
  Kawase, Friend, Shimoda, Inbasekaran, Wu, and Woo}}]{Sirringhaus00}
\bibinfo{author}{\bibfnamefont{H.}~\bibnamefont{Sirringhaus}},
  \bibinfo{author}{\bibfnamefont{T.}~\bibnamefont{Kawase}},
  \bibinfo{author}{\bibfnamefont{R.~H.} \bibnamefont{Friend}},
  \bibinfo{author}{\bibfnamefont{T.}~\bibnamefont{Shimoda}},
  \bibinfo{author}{\bibfnamefont{M.}~\bibnamefont{Inbasekaran}},
  \bibinfo{author}{\bibfnamefont{W.}~\bibnamefont{Wu}}, \bibnamefont{and}
  \bibinfo{author}{\bibfnamefont{E.~P.} \bibnamefont{Woo}},
  \bibinfo{journal}{Science} \textbf{\bibinfo{volume}{290}},
  \bibinfo{pages}{2123} (\bibinfo{year}{2000}).

\bibitem[{\citenamefont{Kawase et~al.}(2001)\citenamefont{Kawase, Sirringhaus,
  Friend, and Shimoda}}]{Kawase01}
\bibinfo{author}{\bibfnamefont{T.}~\bibnamefont{Kawase}},
  \bibinfo{author}{\bibfnamefont{H.}~\bibnamefont{Sirringhaus}},
  \bibinfo{author}{\bibfnamefont{R.~H.} \bibnamefont{Friend}},
  \bibnamefont{and} \bibinfo{author}{\bibfnamefont{T.}~\bibnamefont{Shimoda}},
  \bibinfo{journal}{Adv. Mater.} \textbf{\bibinfo{volume}{13}},
  \bibinfo{pages}{1601} (\bibinfo{year}{2001}).

\bibitem[{\citenamefont{Lehn}(1995)}]{Lehn95}
\bibinfo{author}{\bibfnamefont{J.-M.} \bibnamefont{Lehn}},
  \emph{\bibinfo{title}{Supramolecular Chemistry}} (\bibinfo{publisher}{VCH},
  \bibinfo{address}{Weinheim, Germany}, \bibinfo{year}{1995}).

\bibitem[{\citenamefont{van~de Craats et~al.}(1999)\citenamefont{van~de Craats,
  Warman, Fechtenk\"{o}tter, Brand, Harbison, and M\"{u}llen}}]{Craats99}
\bibinfo{author}{\bibfnamefont{A.~M.} \bibnamefont{van~de Craats}},
  \bibinfo{author}{\bibfnamefont{J.~M.} \bibnamefont{Warman}},
  \bibinfo{author}{\bibfnamefont{A.}~\bibnamefont{Fechtenk\"{o}tter}},
  \bibinfo{author}{\bibfnamefont{J.~D.} \bibnamefont{Brand}},
  \bibinfo{author}{\bibfnamefont{M.~A.} \bibnamefont{Harbison}},
  \bibnamefont{and}
  \bibinfo{author}{\bibfnamefont{K.}~\bibnamefont{M\"{u}llen}},
  \bibinfo{journal}{Adv. Mater.} \textbf{\bibinfo{volume}{11}},
  \bibinfo{pages}{1469} (\bibinfo{year}{1999}).

\bibitem[{\citenamefont{Schenning et~al.}(2001)\citenamefont{Schenning,
  Jonkheijm, Peeters, and Meijer}}]{Schenning01}
\bibinfo{author}{\bibfnamefont{A.~P. H.~J.} \bibnamefont{Schenning}},
  \bibinfo{author}{\bibfnamefont{P.}~\bibnamefont{Jonkheijm}},
  \bibinfo{author}{\bibfnamefont{E.}~\bibnamefont{Peeters}}, \bibnamefont{and}
  \bibinfo{author}{\bibfnamefont{E.~W.} \bibnamefont{Meijer}},
  \bibinfo{journal}{J. Am. Chem. Soc.} \textbf{\bibinfo{volume}{123}},
  \bibinfo{pages}{409} (\bibinfo{year}{2001}).

\bibitem[{\citenamefont{Meijer and Schenning}(2002)}]{Meijer02}
\bibinfo{author}{\bibfnamefont{E.~W.} \bibnamefont{Meijer}} \bibnamefont{and}
  \bibinfo{author}{\bibfnamefont{A.~P. H.~J.} \bibnamefont{Schenning}},
  \bibinfo{journal}{Nature} \textbf{\bibinfo{volume}{419}},
  \bibinfo{pages}{353} (\bibinfo{year}{2002}).

\bibitem[{\citenamefont{Nguyen et~al.}(2000{\natexlab{a}})\citenamefont{Nguyen,
  Wu, Doan, Schwartz, and Tolbert}}]{Nguyen00a}
\bibinfo{author}{\bibfnamefont{T.-Q.} \bibnamefont{Nguyen}},
  \bibinfo{author}{\bibfnamefont{J.}~\bibnamefont{Wu}},
  \bibinfo{author}{\bibfnamefont{V.}~\bibnamefont{Doan}},
  \bibinfo{author}{\bibfnamefont{B.~J.} \bibnamefont{Schwartz}},
  \bibnamefont{and} \bibinfo{author}{\bibfnamefont{S.~H.}
  \bibnamefont{Tolbert}}, \bibinfo{journal}{Science}
  \textbf{\bibinfo{volume}{288}}, \bibinfo{pages}{652}
  (\bibinfo{year}{2000}{\natexlab{a}}).

\bibitem[{\citenamefont{Beljonne et~al.}(2002)\citenamefont{Beljonne, Pourtois,
  Silva, Hennebicq, Herz, Friend, Scholes, Setayesh, M\"{u}llen, and
  Br\'{e}das}}]{Beljonne02}
\bibinfo{author}{\bibfnamefont{D.}~\bibnamefont{Beljonne}},
  \bibinfo{author}{\bibfnamefont{G.}~\bibnamefont{Pourtois}},
  \bibinfo{author}{\bibfnamefont{C.}~\bibnamefont{Silva}},
  \bibinfo{author}{\bibfnamefont{E.}~\bibnamefont{Hennebicq}},
  \bibinfo{author}{\bibfnamefont{L.~M.} \bibnamefont{Herz}},
  \bibinfo{author}{\bibfnamefont{R.~H.} \bibnamefont{Friend}},
  \bibinfo{author}{\bibfnamefont{G.~D.} \bibnamefont{Scholes}},
  \bibinfo{author}{\bibfnamefont{S.}~\bibnamefont{Setayesh}},
  \bibinfo{author}{\bibfnamefont{K.}~\bibnamefont{M\"{u}llen}},
  \bibnamefont{and} \bibinfo{author}{\bibfnamefont{J.~L.}
  \bibnamefont{Br\'{e}das}}, \bibinfo{journal}{Proc. Nat. Acad. Sci. USA}
  \textbf{\bibinfo{volume}{99}}, \bibinfo{pages}{10982} (\bibinfo{year}{2002}).

\bibitem[{\citenamefont{Silva et~al.}(2002)\citenamefont{Silva, Russell, Dhoot,
  Herz, Daniel, Greeham, Arias, Setayesh, M\"{u}llen, and Friend}}]{silva02}
\bibinfo{author}{\bibfnamefont{C.}~\bibnamefont{Silva}},
  \bibinfo{author}{\bibfnamefont{D.~M.} \bibnamefont{Russell}},
  \bibinfo{author}{\bibfnamefont{A.~S.} \bibnamefont{Dhoot}},
  \bibinfo{author}{\bibfnamefont{L.~M.} \bibnamefont{Herz}},
  \bibinfo{author}{\bibfnamefont{C.}~\bibnamefont{Daniel}},
  \bibinfo{author}{\bibfnamefont{N.~C.} \bibnamefont{Greeham}},
  \bibinfo{author}{\bibfnamefont{A.~C.} \bibnamefont{Arias}},
  \bibinfo{author}{\bibfnamefont{S.}~\bibnamefont{Setayesh}},
  \bibinfo{author}{\bibfnamefont{K.}~\bibnamefont{M\"{u}llen}},
  \bibnamefont{and} \bibinfo{author}{\bibfnamefont{R.~H.}
  \bibnamefont{Friend}}, \bibinfo{journal}{J. Phys.: Condens. Matter}
  (\bibinfo{year}{2002}).

\bibitem[{\citenamefont{Herz et~al.}(accepted)\citenamefont{Herz, Daniel,
  Silva, Hoeben, Schenning, Meijer, Friend, and Phillips}}]{Herz03a}
\bibinfo{author}{\bibfnamefont{L.~M.} \bibnamefont{Herz}},
  \bibinfo{author}{\bibfnamefont{C.}~\bibnamefont{Daniel}},
  \bibinfo{author}{\bibfnamefont{C.}~\bibnamefont{Silva}},
  \bibinfo{author}{\bibfnamefont{F.~J.~M.} \bibnamefont{Hoeben}},
  \bibinfo{author}{\bibfnamefont{A.~P. H.~J.} \bibnamefont{Schenning}},
  \bibinfo{author}{\bibfnamefont{E.~W.} \bibnamefont{Meijer}},
  \bibinfo{author}{\bibfnamefont{R.~H.} \bibnamefont{Friend}},
  \bibnamefont{and} \bibinfo{author}{\bibfnamefont{R.~T.}
  \bibnamefont{Phillips}}, \bibinfo{journal}{Phys. Rev. B}
  (\bibinfo{year}{accepted}).

\bibitem[{\citenamefont{Kepler et~al.}(1996)\citenamefont{Kepler, Valencia,
  Jacobs, and McNamara}}]{Kepler96}
\bibinfo{author}{\bibfnamefont{R.~G.} \bibnamefont{Kepler}},
  \bibinfo{author}{\bibfnamefont{V.~S.} \bibnamefont{Valencia}},
  \bibinfo{author}{\bibfnamefont{S.~J.} \bibnamefont{Jacobs}},
  \bibnamefont{and} \bibinfo{author}{\bibfnamefont{J.~J.}
  \bibnamefont{McNamara}}, \bibinfo{journal}{Synth. Met.}
  \textbf{\bibinfo{volume}{78}}, \bibinfo{pages}{227} (\bibinfo{year}{1996}).

\bibitem[{\citenamefont{Klimov et~al.}(1997)\citenamefont{Klimov, McBranch,
  Barashkov, and Ferraris}}]{Klimov97}
\bibinfo{author}{\bibfnamefont{V.~I.} \bibnamefont{Klimov}},
  \bibinfo{author}{\bibfnamefont{D.~W.} \bibnamefont{McBranch}},
  \bibinfo{author}{\bibfnamefont{N.~N.} \bibnamefont{Barashkov}},
  \bibnamefont{and} \bibinfo{author}{\bibfnamefont{J.~P.}
  \bibnamefont{Ferraris}}, \bibinfo{journal}{Chem. Phys. Lett.}
  \textbf{\bibinfo{volume}{277}}, \bibinfo{pages}{109} (\bibinfo{year}{1997}).

\bibitem[{\citenamefont{Vacar et~al.}(1997)\citenamefont{Vacar, Maniloff,
  McBranch, and Heeger}}]{Vacar97}
\bibinfo{author}{\bibfnamefont{D.}~\bibnamefont{Vacar}},
  \bibinfo{author}{\bibfnamefont{E.~S.} \bibnamefont{Maniloff}},
  \bibinfo{author}{\bibfnamefont{D.~W.} \bibnamefont{McBranch}},
  \bibnamefont{and} \bibinfo{author}{\bibfnamefont{A.~J.}
  \bibnamefont{Heeger}}, \bibinfo{journal}{Phys. Rev. B}
  \textbf{\bibinfo{volume}{56}}, \bibinfo{pages}{4573} (\bibinfo{year}{1997}).

\bibitem[{\citenamefont{Maniloff et~al.}(1997)\citenamefont{Maniloff, Klimov,
  and McBranch}}]{Maniloff97}
\bibinfo{author}{\bibfnamefont{E.~S.} \bibnamefont{Maniloff}},
  \bibinfo{author}{\bibfnamefont{V.~I.} \bibnamefont{Klimov}},
  \bibnamefont{and} \bibinfo{author}{\bibfnamefont{D.~W.}
  \bibnamefont{McBranch}}, \bibinfo{journal}{Phys. Rev. B}
  \textbf{\bibinfo{volume}{56}}, \bibinfo{pages}{1876} (\bibinfo{year}{1997}).

\bibitem[{\citenamefont{Dogariu et~al.}(1998)\citenamefont{Dogariu, Vacar, and
  Heeger}}]{Dogariu98}
\bibinfo{author}{\bibfnamefont{A.}~\bibnamefont{Dogariu}},
  \bibinfo{author}{\bibfnamefont{D.}~\bibnamefont{Vacar}}, \bibnamefont{and}
  \bibinfo{author}{\bibfnamefont{A.~J.} \bibnamefont{Heeger}},
  \bibinfo{journal}{Phys. Rev. B} \textbf{\bibinfo{volume}{58}},
  \bibinfo{pages}{10218} (\bibinfo{year}{1998}).

\bibitem[{\citenamefont{Denton et~al.}(1999)\citenamefont{Denton, Tessler,
  Stevens, and Friend}}]{Denton99}
\bibinfo{author}{\bibfnamefont{G.~J.} \bibnamefont{Denton}},
  \bibinfo{author}{\bibfnamefont{N.}~\bibnamefont{Tessler}},
  \bibinfo{author}{\bibfnamefont{M.~A.} \bibnamefont{Stevens}},
  \bibnamefont{and} \bibinfo{author}{\bibfnamefont{R.~H.}
  \bibnamefont{Friend}}, \bibinfo{journal}{Synth. Met.}
  \textbf{\bibinfo{volume}{102}}, \bibinfo{pages}{1008} (\bibinfo{year}{1999}).

\bibitem[{\citenamefont{Nguyen et~al.}(2000{\natexlab{b}})\citenamefont{Nguyen,
  Martini, Liu, and Schwartz}}]{Nguyen00}
\bibinfo{author}{\bibfnamefont{T.~Q.} \bibnamefont{Nguyen}},
  \bibinfo{author}{\bibfnamefont{I.~B.} \bibnamefont{Martini}},
  \bibinfo{author}{\bibfnamefont{J.}~\bibnamefont{Liu}}, \bibnamefont{and}
  \bibinfo{author}{\bibfnamefont{B.~J.} \bibnamefont{Schwartz}},
  \bibinfo{journal}{J. Phys. Chem. B} \textbf{\bibinfo{volume}{104}},
  \bibinfo{pages}{237} (\bibinfo{year}{2000}{\natexlab{b}}).

\bibitem[{\citenamefont{Stevens et~al.}(2001)\citenamefont{Stevens, Silva,
  Russell, and Friend}}]{Stevens01}
\bibinfo{author}{\bibfnamefont{M.~A.} \bibnamefont{Stevens}},
  \bibinfo{author}{\bibfnamefont{C.}~\bibnamefont{Silva}},
  \bibinfo{author}{\bibfnamefont{D.~M.} \bibnamefont{Russell}},
  \bibnamefont{and} \bibinfo{author}{\bibfnamefont{R.~H.}
  \bibnamefont{Friend}}, \bibinfo{journal}{Phys. Rev. B}
  \textbf{\bibinfo{volume}{63}}, \bibinfo{pages}{165213}
  (\bibinfo{year}{2001}).

\bibitem[{\citenamefont{Shimizu et~al.}(2001)\citenamefont{Shimizu, Suto,
  Yamamoto, and Goto}}]{Shimizu01}
\bibinfo{author}{\bibfnamefont{M.}~\bibnamefont{Shimizu}},
  \bibinfo{author}{\bibfnamefont{S.}~\bibnamefont{Suto}},
  \bibinfo{author}{\bibfnamefont{A.}~\bibnamefont{Yamamoto}}, \bibnamefont{and}
  \bibinfo{author}{\bibfnamefont{T.}~\bibnamefont{Goto}},
  \bibinfo{journal}{Phys. Rev. B} \textbf{\bibinfo{volume}{64}},
  \bibinfo{pages}{115417} (\bibinfo{year}{2001}).

\bibitem[{\citenamefont{Silva et~al.}(2001)\citenamefont{Silva, Dhoot, Russell,
  Stevens, Arias, MacKenzie, Greenham, Setayesh, M\"{u}llen, and
  Friend}}]{Silva01}
\bibinfo{author}{\bibfnamefont{C.}~\bibnamefont{Silva}},
  \bibinfo{author}{\bibfnamefont{A.~S.} \bibnamefont{Dhoot}},
  \bibinfo{author}{\bibfnamefont{D.~M.} \bibnamefont{Russell}},
  \bibinfo{author}{\bibfnamefont{M.~A.} \bibnamefont{Stevens}},
  \bibinfo{author}{\bibfnamefont{A.~C.} \bibnamefont{Arias}},
  \bibinfo{author}{\bibfnamefont{J.~D.} \bibnamefont{MacKenzie}},
  \bibinfo{author}{\bibfnamefont{N.~C.} \bibnamefont{Greenham}},
  \bibinfo{author}{\bibfnamefont{S.}~\bibnamefont{Setayesh}},
  \bibinfo{author}{\bibfnamefont{K.}~\bibnamefont{M\"{u}llen}},
  \bibnamefont{and} \bibinfo{author}{\bibfnamefont{R.~H.}
  \bibnamefont{Friend}}, \bibinfo{journal}{Phys. Rev. B.}
  \textbf{\bibinfo{volume}{64}}, \bibinfo{pages}{125211}
  (\bibinfo{year}{2001}).

\bibitem[{\citenamefont{Gadermaier and Lanzani}(2002)}]{Gadermaier02b}
\bibinfo{author}{\bibfnamefont{C.}~\bibnamefont{Gadermaier}} \bibnamefont{and}
  \bibinfo{author}{\bibfnamefont{G.}~\bibnamefont{Lanzani}},
  \bibinfo{journal}{J. Phys.: Condens. Matter} \textbf{\bibinfo{volume}{14}},
  \bibinfo{pages}{9785} (\bibinfo{year}{2002}).

\bibitem[{\citenamefont{Backus et~al.}(1998)\citenamefont{Backus, III, Murnane,
  and Kapteyn}}]{Backus98}
\bibinfo{author}{\bibfnamefont{S.}~\bibnamefont{Backus}},
  \bibinfo{author}{\bibfnamefont{C.~G.~D.} \bibnamefont{III}},
  \bibinfo{author}{\bibfnamefont{M.~M.} \bibnamefont{Murnane}},
  \bibnamefont{and} \bibinfo{author}{\bibfnamefont{H.~C.}
  \bibnamefont{Kapteyn}}, \bibinfo{journal}{Rev. Sci. Instrum.}
  \textbf{\bibinfo{volume}{69}}, \bibinfo{pages}{1207} (\bibinfo{year}{1998}).

\bibitem[{\citenamefont{Jonkheijm et~al.}(unpublished
  data)\citenamefont{Jonkheijm, Hoeben, Schenning, and Meijer}}]{Jonkheijm03}
\bibinfo{author}{\bibfnamefont{P.}~\bibnamefont{Jonkheijm}},
  \bibinfo{author}{\bibfnamefont{F.~J.~M.} \bibnamefont{Hoeben}},
  \bibinfo{author}{\bibfnamefont{A.~P. H.~J.} \bibnamefont{Schenning}},
  \bibnamefont{and} \bibinfo{author}{\bibfnamefont{E.~W.} \bibnamefont{Meijer}}
  (\bibinfo{year}{unpublished data}).

\bibitem[{\citenamefont{Beljonne}(unpublished data)}]{BeljonneC}
\bibinfo{author}{\bibfnamefont{D.}~\bibnamefont{Beljonne}}
  (\bibinfo{year}{unpublished data}).

\bibitem[{\citenamefont{Klimov et~al.}(1998)\citenamefont{Klimov, McBranch,
  Barashkov, and Ferraris}}]{Klimov98}
\bibinfo{author}{\bibfnamefont{V.~I.} \bibnamefont{Klimov}},
  \bibinfo{author}{\bibfnamefont{D.~W.} \bibnamefont{McBranch}},
  \bibinfo{author}{\bibfnamefont{N.}~\bibnamefont{Barashkov}},
  \bibnamefont{and} \bibinfo{author}{\bibfnamefont{J.}~\bibnamefont{Ferraris}},
  \bibinfo{journal}{Phys. Rev. B} \textbf{\bibinfo{volume}{58}},
  \bibinfo{pages}{7654} (\bibinfo{year}{1998}).

\bibitem[{\citenamefont{McBranch et~al.}(1999)\citenamefont{McBranch, Kraabel,
  Xu, Kohlman, Klimov, Bradley, Hsieh, and Rubner}}]{McBranch99}
\bibinfo{author}{\bibfnamefont{D.~W.} \bibnamefont{McBranch}},
  \bibinfo{author}{\bibfnamefont{J.}~\bibnamefont{Kraabel}},
  \bibinfo{author}{\bibfnamefont{S.}~\bibnamefont{Xu}},
  \bibinfo{author}{\bibfnamefont{R.~S.} \bibnamefont{Kohlman}},
  \bibinfo{author}{\bibfnamefont{V.~I.} \bibnamefont{Klimov}},
  \bibinfo{author}{\bibfnamefont{D.~D.~C.} \bibnamefont{Bradley}},
  \bibinfo{author}{\bibfnamefont{B.~R.} \bibnamefont{Hsieh}}, \bibnamefont{and}
  \bibinfo{author}{\bibfnamefont{M.}~\bibnamefont{Rubner}},
  \bibinfo{journal}{Synthetic Metals} \textbf{\bibinfo{volume}{101}},
  \bibinfo{pages}{291} (\bibinfo{year}{1999}).

\bibitem[{\citenamefont{Kraabel et~al.}(2000)\citenamefont{Kraabel, Klimov,
  K\"{o}hlman, Xu, Wang, and McBranch}}]{Kraabel00}
\bibinfo{author}{\bibfnamefont{B.}~\bibnamefont{Kraabel}},
  \bibinfo{author}{\bibfnamefont{V.~I.} \bibnamefont{Klimov}},
  \bibinfo{author}{\bibfnamefont{R.}~\bibnamefont{K\"{o}hlman}},
  \bibinfo{author}{\bibfnamefont{S.}~\bibnamefont{Xu}},
  \bibinfo{author}{\bibfnamefont{H.-L.} \bibnamefont{Wang}}, \bibnamefont{and}
  \bibinfo{author}{\bibfnamefont{D.~W.} \bibnamefont{McBranch}},
  \bibinfo{journal}{Phys. Rev. B} \textbf{\bibinfo{volume}{61}},
  \bibinfo{pages}{8501} (\bibinfo{year}{2000}).

\bibitem[{\citenamefont{Powell and Soos}(1975)}]{Powell75}
\bibinfo{author}{\bibfnamefont{R.~C.} \bibnamefont{Powell}} \bibnamefont{and}
  \bibinfo{author}{\bibfnamefont{Z.~G.} \bibnamefont{Soos}},
  \bibinfo{journal}{J. Lumin.} \textbf{\bibinfo{volume}{11}},
  \bibinfo{pages}{1} (\bibinfo{year}{1975}).

\bibitem[{\citenamefont{F\"{o}rster}(1949)}]{Forster49}
\bibinfo{author}{\bibfnamefont{T.}~\bibnamefont{F\"{o}rster}},
  \bibinfo{journal}{Zeitschrift fur Naturforschung}
  \textbf{\bibinfo{volume}{4a}}, \bibinfo{pages}{321} (\bibinfo{year}{1949}).

\bibitem[{\citenamefont{Eisenthal and Siegel}(1964)}]{Eisenthal64}
\bibinfo{author}{\bibfnamefont{K.~B.} \bibnamefont{Eisenthal}}
  \bibnamefont{and} \bibinfo{author}{\bibfnamefont{S.}~\bibnamefont{Siegel}},
  \bibinfo{journal}{J. Chem. Phys.} \textbf{\bibinfo{volume}{41}},
  \bibinfo{pages}{652} (\bibinfo{year}{1964}).

\bibitem[{\citenamefont{Balagurov and Vaks}(1974)}]{Balagurov74}
\bibinfo{author}{\bibfnamefont{B.~Y.} \bibnamefont{Balagurov}}
  \bibnamefont{and} \bibinfo{author}{\bibfnamefont{V.~G.} \bibnamefont{Vaks}},
  \bibinfo{journal}{Sov. Phys. JETP} \textbf{\bibinfo{volume}{38}},
  \bibinfo{pages}{968} (\bibinfo{year}{1974}).

\bibitem[{\citenamefont{Movaghar et~al.}(1986)\citenamefont{Movaghar,
  Gr\"{u}newald, Ries, B\"{a}ssler, and W\"{u}rtz}}]{Movaghar86}
\bibinfo{author}{\bibfnamefont{B.}~\bibnamefont{Movaghar}},
  \bibinfo{author}{\bibfnamefont{M.}~\bibnamefont{Gr\"{u}newald}},
  \bibinfo{author}{\bibfnamefont{B.}~\bibnamefont{Ries}},
  \bibinfo{author}{\bibfnamefont{H.}~\bibnamefont{B\"{a}ssler}},
  \bibnamefont{and}
  \bibinfo{author}{\bibfnamefont{D.}~\bibnamefont{W\"{u}rtz}},
  \bibinfo{journal}{Phys. Rev. B} \textbf{\bibinfo{volume}{33}},
  \bibinfo{pages}{5545} (\bibinfo{year}{1986}).

\bibitem[{\citenamefont{Mollay et~al.}(1994)\citenamefont{Mollay, Lemmer,
  Kersting, Mahrt, Kurz, Kauffmann, and B\"{a}ssler}}]{Mollay94}
\bibinfo{author}{\bibfnamefont{B.}~\bibnamefont{Mollay}},
  \bibinfo{author}{\bibfnamefont{U.}~\bibnamefont{Lemmer}},
  \bibinfo{author}{\bibfnamefont{R.}~\bibnamefont{Kersting}},
  \bibinfo{author}{\bibfnamefont{R.~F.} \bibnamefont{Mahrt}},
  \bibinfo{author}{\bibfnamefont{H.}~\bibnamefont{Kurz}},
  \bibinfo{author}{\bibfnamefont{H.~F.} \bibnamefont{Kauffmann}},
  \bibnamefont{and}
  \bibinfo{author}{\bibfnamefont{H.}~\bibnamefont{B\"{a}ssler}},
  \bibinfo{journal}{Phys. Rev. B} \textbf{\bibinfo{volume}{50}},
  \bibinfo{pages}{10769} (\bibinfo{year}{1994}).

\bibitem[{\citenamefont{Yan et~al.}(1994)\citenamefont{Yan, Rothberg,
  Papadimitrakopoulos, Galvin, and Miller}}]{Yan94}
\bibinfo{author}{\bibfnamefont{M.}~\bibnamefont{Yan}},
  \bibinfo{author}{\bibfnamefont{L.~J.} \bibnamefont{Rothberg}},
  \bibinfo{author}{\bibfnamefont{F.}~\bibnamefont{Papadimitrakopoulos}},
  \bibinfo{author}{\bibfnamefont{M.~E.} \bibnamefont{Galvin}},
  \bibnamefont{and} \bibinfo{author}{\bibfnamefont{T.~M.}
  \bibnamefont{Miller}}, \bibinfo{journal}{Phys. Rev. Lett.}
  \textbf{\bibinfo{volume}{73}}, \bibinfo{pages}{744} (\bibinfo{year}{1994}).

\bibitem[{\citenamefont{Herz and Phillips}(2000)}]{Herz00}
\bibinfo{author}{\bibfnamefont{L.~M.} \bibnamefont{Herz}} \bibnamefont{and}
  \bibinfo{author}{\bibfnamefont{R.~T.} \bibnamefont{Phillips}},
  \bibinfo{journal}{Phys. Rev. B} \textbf{\bibinfo{volume}{61}},
  \bibinfo{pages}{13691} (\bibinfo{year}{2000}).

\bibitem[{\citenamefont{Cerullo et~al.}(1998)\citenamefont{Cerullo, Nisoli,
  Stagira, Silvestri, Lanzani, Graupner, List, and Leising}}]{Cerullo98}
\bibinfo{author}{\bibfnamefont{G.}~\bibnamefont{Cerullo}},
  \bibinfo{author}{\bibfnamefont{M.}~\bibnamefont{Nisoli}},
  \bibinfo{author}{\bibfnamefont{S.}~\bibnamefont{Stagira}},
  \bibinfo{author}{\bibfnamefont{S.~D.} \bibnamefont{Silvestri}},
  \bibinfo{author}{\bibfnamefont{G.}~\bibnamefont{Lanzani}},
  \bibinfo{author}{\bibfnamefont{W.}~\bibnamefont{Graupner}},
  \bibinfo{author}{\bibfnamefont{E.}~\bibnamefont{List}}, \bibnamefont{and}
  \bibinfo{author}{\bibfnamefont{G.}~\bibnamefont{Leising}},
  \bibinfo{journal}{Chem. Phys. Lett.} \textbf{\bibinfo{volume}{288}},
  \bibinfo{pages}{561} (\bibinfo{year}{1998}).

\bibitem[{\citenamefont{Suna}(1970)}]{Suna70}
\bibinfo{author}{\bibfnamefont{A.}~\bibnamefont{Suna}}, \bibinfo{journal}{Phys.
  Rev. B} \textbf{\bibinfo{volume}{1}}, \bibinfo{pages}{1716}
  (\bibinfo{year}{1970}).

\bibitem[{\citenamefont{Hoeben et~al.}(unpublished data)\citenamefont{Hoeben,
  Herz, Daniel, Jonkheijm, Schenning, Silva, Meskers, Phillips, Friend, and
  Meijer}}]{Hoeben03}
\bibinfo{author}{\bibfnamefont{F.~J.~M.} \bibnamefont{Hoeben}},
  \bibinfo{author}{\bibfnamefont{L.~M.} \bibnamefont{Herz}},
  \bibinfo{author}{\bibfnamefont{C.}~\bibnamefont{Daniel}},
  \bibinfo{author}{\bibfnamefont{P.}~\bibnamefont{Jonkheijm}},
  \bibinfo{author}{\bibfnamefont{A.~P. H.~J.} \bibnamefont{Schenning}},
  \bibinfo{author}{\bibfnamefont{C.}~\bibnamefont{Silva}},
  \bibinfo{author}{\bibfnamefont{S.~C.~J.} \bibnamefont{Meskers}},
  \bibinfo{author}{\bibfnamefont{R.~T.} \bibnamefont{Phillips}},
  \bibinfo{author}{\bibfnamefont{R.~H.} \bibnamefont{Friend}},
  \bibnamefont{and} \bibinfo{author}{\bibfnamefont{E.~W.} \bibnamefont{Meijer}}
  (\bibinfo{year}{unpublished data}).

\bibitem[{\citenamefont{Spano}(2002)}]{Spano02}
\bibinfo{author}{\bibfnamefont{F.~C.} \bibnamefont{Spano}},
  \bibinfo{journal}{J. Chem. Phys.} \textbf{\bibinfo{volume}{116}},
  \bibinfo{pages}{5877} (\bibinfo{year}{2002}).

\bibitem[{\citenamefont{Scholes et~al.}(2001)\citenamefont{Scholes, Jordanides,
  and Fleming}}]{Scholes01}
\bibinfo{author}{\bibfnamefont{G.~D.} \bibnamefont{Scholes}},
  \bibinfo{author}{\bibfnamefont{X.~J.} \bibnamefont{Jordanides}},
  \bibnamefont{and} \bibinfo{author}{\bibfnamefont{G.~R.}
  \bibnamefont{Fleming}}, \bibinfo{journal}{J. Phys. Chem. B}
  \textbf{\bibinfo{volume}{105}}, \bibinfo{pages}{1640} (\bibinfo{year}{2001}).

\bibitem[{\citenamefont{Scholes}(2002)}]{Scholes02}
\bibinfo{author}{\bibfnamefont{G.~D.} \bibnamefont{Scholes}},
  \bibinfo{journal}{Chem. Phys.} \textbf{\bibinfo{volume}{275}},
  \bibinfo{pages}{373} (\bibinfo{year}{2002}).

\end{thebibliography}

\end{document}